\documentclass[aps,preprintnumbers, superscriptaddress, amsmath, showpacs]{revtex4}
\usepackage{epsfig}
\usepackage{psfrag}

\usepackage{graphicx}
\usepackage{dcolumn}
\usepackage{bm}

\begin{document}

\preprint{RU06-3-B}

\title{The imaginary part of the gap function in color superconductivity }

\author{Bo  Feng}
 \email{fengbo@iopp. ccnu. edu. cn}
\affiliation{Institute of Particle Physics,  Huazhong Normal University,
Wuhan,  430079,  China}
\author{ De-fu  Hou}%
 \email{hdf@iopp. ccnu. edu. cn}
 \affiliation{Institute of Particle Physics,  Huazhong Normal University,
Wuhan,  430079,  China}
\author{Jia-rong Li}
 \email{ljr@iopp. ccnu. edu. cn}
\affiliation{Institute of Particle Physics,  Huazhong Normal University,
Wuhan,  430079,  China}
\author{Hai-cang Ren}
\email{ren@summit. rockefeller. edu}
\affiliation{Physics Department,  The Rockefeller University,
1230 York Avenue,  New York,  NY 10021-6399}
\affiliation{Institute of Particle Physics,  Huazhong Normal University,
Wuhan,  430079,  China}

\date{\today}

\begin{abstract}
We clarify general properties of the energy gap regarding its
functional dependence on the energy-momentum dictated
by the invariance under a space inversion or a time reversal. 
Then we derive perturbatively the equation of the imaginary part of the
gap function for dense QCD in weak coupling and generalize our
results from 2SC case to CFL case. We confirm that
the imaginary part is down by $g$ relative to the real part in
weak coupling. The numerical results show that,
up to the leading order, the imaginary part is no larger than one
MeV at extremely large densities and can be as large as several
MeV for the densities  of physical interest. 
\end{abstract}

\maketitle

\section{Introduction}\label{sec:level1}

At sufficiently high density a quark matter is expected to become
a color superconductor,  which has stirred a lot of interests
 \cite{KF,dk}.  The mechanism of color superconductivity is essentially the quark analogue of BCS
scenario\cite{JLJ},  which implies that if there is an attractive
interaction in a cold Fermi sea,  the system is unstable against
 the formation of a diquark(Cooper pair)
condensate\cite{B,DA,ARW1}. In QCD case at asymptotically high
density,  the dominant interaction between two quarks is due to
the one-gluon exchange, which is attractive in the color
antitriplet channel.  A first principle calculation can be carried
out perturbatively at this density and the asymptotic formula for
the energy gap and the transition temperature have been derived
\cite{D, TF, RD, RD2, HMSW, C, WJH1, WJH2, QDH}.  At moderately
high density that is accessible in a compact star,  the instanton
mediated interaction may dominate, which is again attrarctive in
the color antisymmetric channel. Because of its nonperturbative
origin,  one has to resort to various phenomenological models,
such as NJL model\cite{YG, JK, RTEM, M} to explore the color
superconductivity in this case.

Color superconductivity has a very rich phase structure because of
its color and flavor as well as spin degrees of
freedom\cite{KF}. It is well-established that at extremely high
density,  where the chemical potential is much larger than the
strange quark mass,  the ground state is characterized by the
so-called color-flavor locked(CFL)condensate\cite{MKF}.  But actually
the chemical potential is unlikely to be much larger than 500Mev
in the cores of a compact star.  Therefore,  the heavier strange
quark may not be able to participate the pairing with up and down
quarks.  The color superconducting phase with only two light quarks
is normally called the 2SC phase.  However, because of the
$\beta$-equilibrium and the charge neutrality conditions of 
the quark system, nontrivial relations will be imposed
between the chemical potentials of different quark flavors\cite{MK}.  In
turn,  such relations could substantially influence the pairing
dynamics.  It was shown that gapless
2SC\cite{IM, MI, MPW}and gapless CFL\cite{MCK}as well as gapless
LOFF\cite{IDH}could appear in quark matter regarding the influence
of $\beta$-equilibrium and the charge neutrality.

One of the main consequences of color superconductivity in dense
quark matter is the opening of an energy gap,  which impacts
on the transport and thermodynamic properties of quark matter.
The general pairing potential is retarded and contains damping
terms (Landau damping in case of one-gluon exchange of QCD),
which corresponds to branch cuts along the real axis of the complex
energy plane.  Consequently,  the gap function depends on the energy-
momentum and acquires a nontrivial
imaginary part along the axis of real energy.
A gap function with a nonzero
imaginary part is actually well known from strongly coupled
electronic superconductors,  as studied in Eliashberg theory\cite{G}.
But the parallel case has not been analyzed in detail until recently
by Rueter \cite{philipp} .
Our main work in this paper is
to derive the equation of the imaginary part of the gap function
in weak coupling.  Our results confirm that the imaginary
part of the gap function is down by $g$ relative to the real part
in weak coupling.  At extremely large densities,  the imaginary part
of the gap function is smaller than one MeV.  For chemical
potential that are of physical interest,  $\mu<$1 GeV,  the
imaginary part can be as large as several MeV.

The current work is organized as follows.  In  section 2, 
 we shall demonstrate  some general properties of the energy gap as a function of the energy and 
 momentum. Then we  review the  Eliashberg equations for the energy gap at real energy in Section 3.  Its solution will be presented in
 Section 4 and the concluding
remarks will then be given in Section 5. Moreover,  technical details
on the derivation of the real part of the  gap  will be sketched in  Appendix A.
 Our units are
$\hbar=c=k_B=1$ and 4-vectors are denoted by capital letters,
$K \equiv K^\mu = (\vec{k},\omega)$ in our formulas.

\section{General properties of the energy gap as a function of energy and momentum}

Before embarking on the calculation of the imaginary part of the
energy gap, we shall clarify some general
properties of the energy gap regarding its functional dependence
on the energy and momentum dictated by the invariance under
a space inversion, ${\cal P}$, or a time reversal, ${\cal T}$.
These properties are model independent and exact to all orders of
the coupling strength. While, some of them may be known in the
case of electronic superconductivity, they remain murky for color
superconductivity. Since a large body of formula employed in this
section stems from that of Ref.\cite{TDLee}, we shall follow the
convention there for the gamma matrices (i.e. all gamma matrices
are hermitian with $\gamma_1$, $\gamma_3$ antisymmetric and 
$\gamma_2$, $\gamma_4$ symmetric).

Starting with the Schroedinger representation, the Nambu-Gorkov(NG) quark field reads
\begin{equation}
\Psi(\vec r,0)=\left(\begin{array}{cc}\psi(\vec r,0)\\ \psi_C(\vec r,0)\end{array}\right)
\end{equation}
\label{NG}
and is self-conjugate, i.e.
\begin{equation}
\Psi(\vec r,0)=C\tilde\Psi^\dagger(\vec r,0),
\label{conjug}
\end{equation}
where $\psi$ is a Dirac spinor, $\psi_c=i\gamma_2\tilde{\psi^\dagger}$ with the tilde standing
for transpose and
\begin{equation}
C=i\left(\begin{array}{cc}0 & \gamma_2 \\ \gamma_2 & 0
\end{array}\right)=\tilde C=-C^\dagger.
\end{equation}
Color and flavor indexes are suppressed. We choose the phase of $\psi$ such that the
transformation rules under ${\cal P}$ and
${\cal T}$ are given by
\begin{equation}
{\cal P}\Psi(\vec r,0){\cal P}^\dagger=U_P\Psi(-\vec r,0),
\label{parity1}
\end{equation}
\begin{equation}
{\cal P}\Psi^\dagger(\vec r,0){\cal P}^\dagger=\Psi^\dagger(-\vec r,0)U_P^\dagger,
\label{parity2}
\end{equation}
\begin{equation}
{\cal T}\Psi(\vec r,0){\cal T}^{-1}=U_T\Psi(\vec r,0),
\label{time1}
\end{equation}
and
\begin{equation}
{\cal T}\Psi^\dagger(\vec r,0){\cal T}^{-1}=\Psi^\dagger(\vec r,0)U_T^\dagger,
\label{time2}
\end{equation}
where $U_P={\rm diag.}(\gamma_4,-\gamma_4)=U_P^\dagger$,
$U_T={\rm diag}(i\eta\gamma_1\gamma_3,i\eta^*\gamma_1\gamma_3)
=-\tilde U_T$
with $\eta$ a phase factor ($|\eta|=1$)
independent of the coordinates and internal indexes.

The self-conjugation (\ref{conjug}) and the transformation rules (\ref{parity1})-(\ref{time2})
in the Heisenberg and Matsubara
representations follow from the real(imaginary) time development formula of an operator
$O$, i.e.
\begin{equation}
O(t) = e^{iHt}O(0)e^{-iHt}
\label{heisenberg}
\end{equation}
and
\begin{equation}
O(\tau) = e^{H\tau}O(0)e^{-H\tau}
\label{matsubara}
\end{equation}
with $H$ the Hamiltonian. Note that the similarity transformation (\ref{matsubara}) is not unitary and thus does
not preserve the relation of hermitian conjugation. In particular, on writing
\begin{equation}
\bar O(\tau)\equiv e^{H\tau}O^\dagger(0)e^{-H\tau}
\end{equation}  
we have $\bar O(\tau)=O^\dagger(-\tau)$
and vice versa. The over bar in this section does not mean a
right multiplication by $\gamma_4$.

\subsection{Real time formulation}

The retarded NG quark propagator is defined by
\begin{equation}
{\cal S}_{\alpha\beta}(\vec r,t)=i<\lbrace\Psi_\alpha(\vec r,t),\Psi_\lambda^\dagger(0,0)\rbrace>
(\Gamma_4)_{\lambda\beta}\theta(t)
\label{retard}
\end{equation}
where $\Gamma_4={\rm diag}(\gamma_4,\gamma_4)$ and $<O>$ denotes the thermal average of the
operator $O$, i.e.
\begin{equation}
<O>\equiv\frac{{\rm Tr}e^{-\beta H}O}{{\rm Tr}e^{-\beta H}}
\end{equation}
with $\beta=1/T$ the inverse temperature.
The Fourier transformation of (\ref{retard}) reads
\begin{equation}
{\cal S}(\vec p,\omega)=\int_{-\infty}^\infty dt\int d^3\vec r
e^{i\omega t-i\vec p\cdot\vec r}{\cal S}(\vec r,t)
\label{fourier}
\end{equation}
with $(\vec p,\omega)$ the momentum and the energy. On writing
\begin{equation}
{\cal S}^{-1}(\vec p,\omega)=\left(\begin{array}{cc}K(\vec p,\omega)\,\,\,   & \Phi(\vec p,\omega)\\
\Phi^\prime(\vec p,\omega) \,\,\,  & K^\prime(\vec p,\omega)\end{array}\right)
\label{relationI}
\end{equation}
The off-diagonal matrix elements carry the long range order and the gap functions for 2SC are
extracted according
\begin{equation}
\Phi(\vec p,\omega)=-\eta^*\phi(\vec p,\omega)\gamma_5\lambda_2\tau_2
\end{equation}
and
\begin{equation}
\Phi^\prime(\vec p,\omega)=\eta\phi^\prime(\vec p,\omega)\gamma_5\lambda_2\tau_2
\end{equation}
with $\lambda_2$ the second Gell-Mann matrix acting on colors and $\tau_2$ the
second Pauli matrix acting on flavors.

The self-conjugate relation (\ref{conjug}) in the Heisenberg representation reads
\begin{equation}
\Psi(\vec r,t)=C\tilde\Psi^\dagger(\vec r,t).
\label{conjugt}
\end{equation}
Taking the complex conjugate of (\ref{retard}) and applying the relation
(\ref{conjugt}) we find
\begin{eqnarray}
{\cal S}_{\alpha\beta}^*(\vec r,t) &=& -iC_{\lambda\rho}(C^\dagger)_{\gamma\alpha}
<|\lbrace\Psi_\gamma(\vec r,t),\Psi_\rho^\dagger(0,0)\rbrace|>(\Gamma_4)_{\lambda\beta}\\ \nonumber
&=& -(C^\dagger)_{\alpha\gamma}{\cal S}_{\gamma\rho}(\vec r,t)
(\Gamma_4\tilde C\Gamma_4)_{\rho\beta}
\end{eqnarray}
Suppressing the indexes and using the relations $\tilde C=C$ and
$\Gamma_4\tilde C\Gamma_4=-C$, we obtain that
\begin{equation}
{\cal S}^*(\vec r,t)=C^\dagger{\cal S}(\vec r,t)C
\label{relationX}
\end{equation}
and its Fourier transform
\begin{equation}
{\cal S}^*(-\vec p,-\omega)=C^\dagger{\cal S}(\vec p,\omega)C.
\label{relationCH}
\end{equation}
The same relation should also apply to the inverse propagator, i.e.
\begin{equation}
{\cal S}^{*-1}(-\vec p,-\omega)=C^\dagger {{\cal S}^{-1}}(\vec p,\omega)C,
\end{equation}
which implies that
\begin{equation}
\phi^\prime(\vec p,\omega)=\phi^*(-\vec p,-\omega).
\label{selfh}
\end{equation}
Therefore, $\Phi(\vec p,\omega)=-\eta^*\phi(\vec p,\omega)\gamma_5\lambda_2\tau_2$
and $\Phi^\prime(\vec p,\omega)=\eta\phi^*(-\vec p,-\omega)\gamma_5\lambda_2\tau_2$
We are left with
only one complex gap function, $\phi(\vec p,\omega)$, to consider.

The ${\cal P}$ and ${\cal T}$ transformation rules of the Heisenberg operators $\Psi(\vec r,t)$
and $\Psi^\dagger(\vec r,t)$ read
\begin{equation}
{\cal P}\Psi(\vec r,t){\cal P}^\dagger=U_P\Psi(-\vec r,t),
\label{parity1t}
\end{equation}
\begin{equation}
{\cal P}\Psi^\dagger(\vec r,t){\cal P}^\dagger=\Psi^\dagger(-\vec r,t)U_P^\dagger
\label{parity2t}
\end{equation}
\begin{equation}
{\cal T}\Psi(\vec r,t){\cal T}^{-1}=U_T\Psi(\vec r,-t),
\label{time1t}
\end{equation}
and
\begin{equation}
{\cal T}\Psi^\dagger(\vec r,t){\cal T}^{-1}=\Psi^\dagger(\vec r,-t)U_T^\dagger.
\label{time2t}
\end{equation}
The invariance of the Hamiltonian of the color-superconductivity implies that
\begin{equation}
<O>=<{\cal P}O{\cal P}^\dagger>
\label{parity}
\end{equation}
and
\begin{equation}
<O>=<{\cal T}O^\dagger{\cal T}^{-1}>.
\label{treversal}
\end{equation}
The proof of Eqs.(\ref{parity}) and (\ref{treversal}) follows from
the observation that under ${\cal P}$ or ${\cal T}$, a state is
either transformed to itself ( e.g. the ground state ) or another
state of the same eigenvalue of $H$. The hermitian conjugation on
RHS of (\ref{treversal}) is because of the anti-unitarity of
${\cal T}$. See \cite{TDLee} for details.

The implication of ${\cal P}$ invariance can be obtained trivially. We have
\begin{equation}
{\cal S}(\vec r,t)=U_P{\cal S}(-\vec r,t)U_P^\dagger
\label{relationPH}
\end{equation}
or equivalently
\begin{equation}
{\cal S}^{-1}(\vec p,\omega)=U_P{\cal S}^{-1}(-\vec p,\omega)U_P^\dagger,
\end{equation}
which gives rise to
\begin{equation}
\phi(-\vec p,\omega)=\phi(\vec p,\omega).
\label{parityh}
\end{equation}
Thus the gap function is even with respect to the spatial
momentum. As to the time reversal, Eq.(\ref{treversal}) together
with Eqs.(\ref{time1t}) and (\ref{time2t}) yield
\begin{eqnarray}
{\cal S}_{\alpha\beta}(\vec r,t) &=&
i<|{\cal T}\lbrace\Psi_\lambda(0,0),\Psi_\alpha^\dagger(\vec r,t)\rbrace{\cal T}^{-1}|>
(\Gamma_4)_{\lambda\beta}\\ \nonumber
&=& i(U_T)_{\lambda\rho}(U_T^\dagger)_{\gamma\alpha}
<|\lbrace\Psi_\rho(0,0),\Psi_\gamma^\dagger(\vec r,-t)\rbrace|>
(\Gamma_4)_{\lambda\beta}
=(U_T^*\Gamma_4)_{\alpha\rho}\tilde{\cal S}_{\rho\lambda}(-\vec r,t)
(\tilde U_T\Gamma_4)_{\lambda\beta}.
\end{eqnarray}
Therefore
\begin{equation}
{\cal S}(\vec r,t)=U_T^*\Gamma_4\tilde{\cal S}(-\vec r,t)\tilde U\Gamma_4
=U_T^\dagger\Gamma_4\tilde{\cal S}(-\vec r,t)U_T\Gamma_4,
\label{relationT}
\end{equation}
which gives rise to the relation
\begin{equation}
{\cal S}^{-1}(\vec p,\omega)=U_T^\dagger\Gamma_4\tilde{\cal S}^{-1}(-\vec p,\omega)U_T\Gamma_4.
\label{reversalT}
\end{equation}
It follows from (\ref{selfh}) and (\ref{reversalT}) that
\begin{equation}
\phi^*(\vec p,-\omega)=\phi(\vec p,\omega).
\end{equation}
Consequently, we arrive at

\noindent
{\it{Theorem 1}}: The invariance of 2SC under both ${\cal P}$ and ${\cal T}$
implies the following off-diagonal structure of the inverse NG propagator
of real energy
(\ref{relationI}):
\begin{eqnarray}
\Phi(\vec p,\omega) &=& -\eta^*\phi(\vec p,\omega)\gamma_5\lambda_2\tau_2\\ \nonumber
\Phi^\prime(\vec p,\omega) &=& \eta\phi(\vec p,\omega)\gamma_5\lambda_2\tau_2
\end{eqnarray}
where $\phi(\vec p,\omega)$ an even function of $\vec p$ and satisfies the relation
$\phi(\vec p,\omega)=\phi^*(\vec p,-\omega)$.

Decomposing $\phi(\vec p,\omega)$ into its real and imaginary
parts, $\phi(\vec p,\omega)={\rm Re}\phi(\vec p,\omega)+i{\rm
Im}\phi(\vec p,\omega)$, we find that ${\rm Re}\phi(\vec p)$(${\rm
Im}\phi(\vec p,\omega)$) is an even(odd) function of $\omega$, a
statement that can also be established by analytically continuating the 
real solution of the gap equation with Matsubara energy, shown in  
Ref. \cite{philipp} and in subsequent sections. But the reality of the 
gap function of Matsubara energy follows from the invariance under 
${\cal P}$ and ${\cal T}$ as is stated in the theorem 2 below.

\subsection{Matsubara formulation}

The Matsubara quark propagator is defined by
\begin{equation}
S_{\alpha\beta}(\vec r,\tau)=[<\Psi_\alpha(\vec r,\tau)\Psi_\lambda^\dagger(0,0)>\theta(\tau)
-<\Psi_\lambda^\dagger(0,0)\Psi_\alpha(\vec r,\tau)>\theta(-\tau)]
(\Gamma_4)_{\lambda\beta}
\label{euclid}
\end{equation}
The absence of the factor $i$ in comparison with (\ref{retard}) is necessary in order to match the
Fourier transformation of (\ref{euclid}),
\begin{equation}
S_\nu(\vec p)=\int_0^\beta d\tau\int d^3\vec r
e^{i\nu\tau-i\vec p\cdot\vec r}S(\vec r,\tau),
\end{equation}
to the analytic continuation of Eq. (\ref{fourier}), i.e.
$S_\nu(\vec p)={\cal S}(\vec p,i\nu)$, where $\nu=2\pi T(n+\frac{1}{2})$ is
the Matsubara energy. On writing
\begin{equation}
S_\nu^{-1}(\vec p)=\left(\begin{array}{cc}K_\nu(\vec p)\,\,\,   & \Phi_\nu(\vec p)\\
\Phi_\nu^\prime(\vec p) \,\,\,  & K_\nu^\prime(\vec p)\end{array}\right)
\label{relationJ}
\end{equation}
with $\Phi_\nu(\vec p)=-\eta^*\phi_\nu(\vec p)\gamma_5\lambda_2\tau_2$ and
$\Phi_\nu^\prime(\vec p)=\eta\phi_\nu^\prime(\vec p)\gamma_5\lambda_2\tau_2$
The self-conjugate relation
\begin{equation}
\Psi(\vec r,\tau)=C\tilde{\bar\Psi}(\vec r,t).
\label{conjugtau}
\end{equation}
implies that
\begin{equation}
S_\nu^{*-1}(-\vec p)=C^\dagger S_\nu^{-1} (\vec p)C,
\label{relationCM}
\end{equation}
which leads to
\begin{equation}
\phi'_\nu(\vec p)=\phi_\nu^*(-\vec p).
\label{selfm}
\end{equation}

It follows from (\ref{parity}) and the ${\cal P}$-transformation of the Matsubara operators
$\Psi(\vec r,\tau)$ and $\bar\Psi(\vec r,\tau)$
\begin{equation}
{\cal P}\Psi(\vec r,\tau){\cal P}^\dagger=U_P\Psi(-\vec r,\tau),
\label{parity1tau}
\end{equation}
and
\begin{equation}
{\cal P}\bar\Psi(\vec r,\tau){\cal P}^\dagger=\bar\Psi(-\vec r,\tau)U_P^\dagger,
\label{parity2tau}
\end{equation}
that
\begin{equation}
S_\nu^{-1}(\vec p)=U_PS_\nu^{-1}(-\vec p)U_P^\dagger,
\end{equation}
which implies that
\begin{equation}
\phi_\nu(\vec p)=\phi_\nu(-\vec p).
\label{paritym}
\end{equation}
Finally, the ${\cal T}$ invariance formula (\ref{treversal}) together with the transformation
rules
\begin{equation}
{\cal T}\Psi(\vec r,\tau){\cal T}^{-1}=U_T\Psi(\vec r,\tau),
\label{time1tau}
\end{equation}
and
\begin{equation}
{\cal T}\bar\Psi(\vec r,\tau){\cal T}^{-1}=\bar\Psi(\vec r,\tau)U_T^\dagger,
\label{time2tau}
\end{equation}
give rise to
\begin{equation}
S_\nu^{-1}(\vec p)=U_T^\dagger\Gamma_4\tilde S_\nu^{-1}(-\vec p)U_T\Gamma_4.
\end{equation}
It follows then that
\begin{equation}
\phi_\nu^*(\vec p)=\phi_\nu(\vec p)
\label{tm}
\end{equation}
Combining (\ref{selfm}, (\ref{paritym}) and (\ref{tm}), we end up with

\noindent
{\it{Theorem 2}}: The invariance of 2SC under both ${\cal P}$ and ${\cal T}$
implies the following off-diagonal structure of the inverse NG propagator
of Matsubara energy
(\ref{relationJ}):
\begin{eqnarray}
\Phi_\nu(\vec p) &=& -\eta^*\phi_\nu(\vec p)\gamma_5\lambda_2\tau_2\\ \nonumber
\Phi_\nu^\prime(\vec p) &=& \eta\phi_\nu(\vec p)\gamma_5\lambda_2\tau_2
\end{eqnarray}
where $\phi_\nu(\vec p)$ a real and even function of $\vec p$.

In the next section, we shall derive the equation for the gap function of real energy,
$\phi(\vec p,\omega)$ by analytic continuation of the gap equation in Matsubara
formulation. To unify the notation throughout the continuation process, we
shall write $\phi(\vec p,i\nu)$ for $\phi_\nu(\vec p)$. Because of the weak coupling
approximation employed, the dependence on $\vec p$ may can ignored, leaving the gap
a function of energy only.

The two theorems we have established for 2SC apply to the the gap function 
of CFL as well.

\section{Eliashberg equation for gap parameter at real energy}

The original Eliashberg theory formulated for an electronic
superconductor of strong pairing force regards both the energy gap
and the quasi particle weight ( wave function renormalization )
analytic functions on the complex energy plane cut along the real
axis.  They are determined {\it{at equal footing}} from a pair of
self consistent equations of the electron self energy with respect
to NG basis,  one refers to the diagonal part and the other to the
off-diagonal part.  The latter one corresponds to the gap equation
in the usual sense.  For QCD at asymptotic quark density, however,
the full complexity of the Eliashberg equations is unnecessary,
and one expects the following weak coupling expansion of the
energy gap function of the real energy $\omega$;
\begin{equation}
\phi(\omega)=e^{-\frac{\kappa}{g}}
\Big[f_0\Big(g\ln\frac{\omega_c}{\omega}\Big)+gf_1\Big(g\ln\frac{\omega_c}{\omega}\Big)
+. . . \Big]
\label{expansion}
\end{equation}
for $g\ll 1$, but  $g\ln\frac{\omega_c}{\omega}=O(1)$,  where $\omega_c=O(\frac{\mu}{g^5})$ \,  and
powers of $\ln g$ are regarded O(1) in the expansion.  Both the exponent $\kappa$ and the
function $f_0$ are known in the literature and are referred to as the leading and subleading
contributions of the gap function. The leading order imaginary part shows up in the function
$f_1$ and is therefore corresponds to the sub-subleading contribution to the complex gap function.

To determine the ${\rm Im}f_1$, we need only one Eliashberg
equation, which is the analytic continuation from the gap equation
of Euclidean energy. For the sake of notation simplicity, we focus
mainly on the 2SC case. The generalization to CFL case is
straightforward and will be addressed at the end of the next
section.

The gap equation of 2SC for either right- or left- handed gap function
with Euclidean momentum is given in Ref.\cite{RD} \setlength\arraycolsep{1pt}
\begin{eqnarray}
\nonumber
\phi(K)&=&\frac{2}{3}g^2\frac{T}{V}\sum\limits_{q_0}\frac{\phi(Q)}{[q_0/Z(q_0)]^2-[\epsilon_q(\phi)]^2}
[\Delta_l(K-Q)\frac{1+\hat{k}\cdot\hat{q}}{2}\\
&+&\Delta_t(K-Q)(-\frac{3-\hat{k}\cdot\hat{q}}{2}+
\frac{1+\hat{k}\cdot\hat{q}}{2}\frac{(k-q)^2}{(\vec{k}-\vec{q})^2})]
\label{gap}
\end{eqnarray}
where $T/V\sum\limits_{q_0}\equiv T\sum\limits_{n}\int
d^3\vec{q}/(2\pi)^3$in the infinite-volume limit, $n$ labels the
Matsubara frequencies $-iq_0\equiv(2n+1)\pi T$, and
\begin{equation}
Z(q_0)\equiv[1+\bar{g}^2\rm{ln}(M^2/|q_{0}|^2)]^{-1}
\label{Zfactor}
\end{equation}
is the quark wave-function renormalization factor\cite{C, WJH1,
WJH2, QDH} in the normal phase. It is worth mentioning here that
this equation does not contain color, flavor and Dirac indices any
more,  since they have been computed explicitly.  For the technical 
details that leads from (\ref{relationJ}) to (\ref{gap}), see Ref.
\cite{RD,RD2}. $\Delta_l, _t$ are the longitudinal and transverse
gluon propagators respectively. We will give their expressions
later. The feedback of the energy gap to $Z(q_0)$, which should be
determined from the other Eliashberg equation has been neglected.

To perform the Matsubara sum over quark energies $q_0$,  we
introduce the spectral representations\cite{RD}. For the gluon
propagators
\begin{eqnarray}
\Delta_l(p)&=&-\frac{1}{p^2}+\int\limits_{0}^{1/T}d\tau e^{p_0\tau}
\Delta_l(\tau, \vec{p})\\
\Delta_t(p)&=&\int\limits_{0}^{1/T}d\tau e^{p_0\tau}
\Delta_t(\tau, \vec{p})\\
\Delta_{l, t}(\tau,p)&\equiv&
\int\limits_{0}^{\infty}d\omega\rho_{l, t}(\omega, \vec{p})\{[1+n_B(\omega/T)]e^{-\omega\tau}+n_B(\omega/T)e^{\omega\tau}\}
\end{eqnarray}
The spectral densities are given by
\begin{equation}
\rho_{l, t}(\omega, \vec{p})=\rho_{l, t}^{pole}(\omega, p)\delta[\omega-\omega_{l, t}(\vec{p})]
+\rho_{l, t}^{cut}(\omega, \vec{p})\theta(p-\omega)
\label{gspectral}
\end{equation}
where, $n_B(x)\equiv1/(e^x-1)$ is the Bose-Einstein distribution
function,  the expressions of $\rho_{l, t}^{pole}$
and$\rho_{l, t}^{cut}$ will be given explicitly later.  We also
introduce a spectral representation for the quantity\cite{RD}
\begin{equation}
\Xi(Q)\equiv\frac{\phi(Q)}{[q_{0}/Z(q_0)]^{2}-\epsilon_q(\phi)^2}\equiv\int\limits_{0}^{1/T}d\tau
e^{q_0\tau}\Xi(\tau, \vec{q})
\label{agreen}
\end{equation}
\begin{equation}
\Xi(\tau, \vec{q})\equiv\int\limits_{0}^{\infty}d\omega\tilde{\rho}(\omega, \vec{q})\{[1-n_F(\omega/T)]e^{-\omega\tau}-n_F(\omega/T)e^{\omega\tau}\}
\end{equation}
Where  $n_F(x)\equiv1/(e^x+1)$ is the Fermi-Dirac distribution
function.  In Ref. \cite{RD},  Pisarski and Rischke
approximated the spetral density $\tilde\rho(\omega, q)$ by a delta
function corresponding to the quasi-particle mass shell,  which is appropriate for a
real gap.  In order to extract the imaginary part of the gap,  this approximation will not be
made here and we shall follow the off-shell formulation of
the conventional Eliashberg theory that takes the energy as the
argument of the gap\cite{TF,G}.

The Matsubara sums over $q_0$ can be computed
as($\vec{p}\equiv\vec{k}-\vec{q}$)
\begin{widetext}
\begin{eqnarray}
\nonumber
T\sum_{q_0}\Delta_l(K-Q)\Xi(Q)&=&\int\limits_{0}^{\infty} d\varepsilon \, \tilde{\rho}(\varepsilon, \vec{q})\{-\frac{1}{2}\rm{tanh}(\frac{\varepsilon}{2T})+\int\limits_{0}^{\infty}d\nu\rho_l(\nu, \vec{p})
\times[\frac{1}{2}\rm{tanh}(\frac{\varepsilon}{2T})
(\frac{1}{k_0+\nu+\varepsilon}-\frac{1}{k_0-\nu-\varepsilon}\\
\nonumber&-&\frac{1}{k_0-\nu+\varepsilon}+\frac{1}{k_0+\nu-\varepsilon})
+\frac{1}{2}\rm{coth}(\frac{\nu}{2T})(\frac{1}{k_0+\nu+\varepsilon}-\frac{1}{k_0-\nu-\varepsilon}+\frac{1}{k_0-\nu+\varepsilon}\\&-&\frac{1}{k_0+\nu-\varepsilon})]\}
\end{eqnarray}
\begin{eqnarray}
\nonumber
T\sum_{q_0}\Delta_t(K-Q)\Xi(Q)&=&\int\limits_{0}^{\infty}d\nu\rho_t(\nu, \vec{p})\int\limits_{0}^{\infty}d\varepsilon\tilde{\rho}
(\varepsilon, \vec{q})\times[\frac{1}{2}\rm{tanh}(\frac{\varepsilon}{2T})
(\frac{1}{k_0+\nu+\varepsilon}-\frac{1}{k_0-\nu-\varepsilon}-\frac{1}{k_0-\nu+\varepsilon}\\
&+&\frac{1}{k_0+\nu-\varepsilon})
+\frac{1}{2}\rm{coth}(\frac{\nu}{2T})(\frac{1}{k_0+\nu+\varepsilon}-\frac{1}{k_0-\nu-\varepsilon}+\frac{1}{k_0-\nu+\varepsilon}-\frac{1}{k_0+\nu-\varepsilon})]
\end{eqnarray}
\end{widetext}
For the first step,  we are only  interested in  zero temperature case.
 Setting $T=0$ leads
\begin{widetext}
\begin{eqnarray}
T\sum_{q_0}\Delta_l(K-Q)\Xi(Q)=\int\limits_{0}^{\infty}d\nu\rho_l(\nu, \vec{p})\int\limits_{0}^{\infty}d\varepsilon\tilde{\rho}
(\varepsilon, \vec{q})\times
(\frac{1}{k_0+\nu+\varepsilon}-\frac{1}{k_0-\nu-\varepsilon})
-\frac{1}{p^2}\int\limits_{0}^{\infty}d\varepsilon\tilde{\rho}(\varepsilon, \vec{q})
\end{eqnarray}
\begin{eqnarray}
T\sum_{q_0}\Delta_t(K-Q)\Xi(Q)=\int\limits_{0}^{\infty}d\nu\rho_t(\nu, \vec{p})\int\limits_{0}^{\infty}d\varepsilon\tilde{\rho}
(\varepsilon, \vec{q})
\times(\frac{1}{k_0+\nu+\varepsilon}-\frac{1}{k_0-\nu-\varepsilon})
\end{eqnarray}
\end{widetext}
Here we take both the energy and momentum  as the arguments of the
imaginary part of the gap function.  At weak coupling limit,  all
quark momenta are close to the fermi surface , we set  $q\simeq
k\simeq\mu$,  the trivial parts in the Eq. (\ref{gap}) can be
simplified as
\begin{equation}
\frac{1+\hat{k}\cdot\hat{q}}{2}=\frac{(k+q)^2-p^2}{4kq}\simeq1
\end{equation}
\begin{equation}
-\frac{3-\hat{k}\cdot\hat{q}}{2}+
\frac{1+\hat{k}\cdot\hat{q}}{2}\frac{(k-q)^2}{(\vec{k}-\vec{q})^2}=
-1-\frac{p^2}{4kq}+\frac{(k^2-q^2)^2}{4kqp^2}\simeq-1
\label{approx2}
\end{equation}
It is now the energy dependence of the spectral densities which
provides the interesting phenomena.  So the integral of $\vec{q}$
can be written as
\begin{equation}
\int d^3\vec{q}=\int q^2dqd\cos\theta d\varphi=\int
q^2dq\frac{p}{k\cdot q}dpd\varphi\simeq\int
dqpdpd\varphi|_{k\simeq q=\mu}
\end{equation}
The best way to extract the imaginary part of the gap equation,  as
is shown in Mahan's book,  is to integrate $q-\mu$ first and then
to integrate the polar angles of $\vec q$. To perform the integral
of $q-\mu$,  we introduce the expression of
$\tilde{\rho}(\varepsilon, \vec{q})$.  The relation between
$\tilde{\rho}(\varepsilon, \vec{q})$ and $\Xi(Q)$ is a special case
of the Kramers-Kronig relation,  when $q_0$ approaches to the real
axis from above.
\begin{equation}
\tilde{\rho}
(\varepsilon, \vec{q})=\frac{1}{\pi}{\rm{Im}}[\Xi(\varepsilon, \vec{q})]
\label{spect-f}
\end{equation}
For a complex  gap function, the spectral density  $\tilde{\rho}(\varepsilon, \vec{q})$  has a finite
width.  Only when $\rm{Im}\phi$ vanishes,  the spectral density comes back to the
on-shell form  that used in Ref.\cite{RD}

\begin{equation}
\int
dqT\sum_{q_0}\Delta_l(K-Q)\Xi(Q)=-\int\limits_{0}^{\infty}d\nu\rho_l(\nu,
\vec{p})\int\limits_{0}^{\infty}d\varepsilon \rm{Re}f(\varepsilon)
\times(\frac{1}{k_0+\nu+\varepsilon}-\frac{1}{k_0-\nu-\varepsilon})
-\frac{1}{p^2}\int\limits_{0}^{\infty}d\varepsilon\rm{Re}f(\varepsilon)
\label{suml2}
\end{equation}
\begin{equation}
-\int
dqT\sum_{q_0}\Delta_t(K-Q)\Xi(Q)=\int\limits_{0}^{\infty}d\nu\rho_t(\nu,
\vec{p})\int\limits_{0}^{\infty}d\varepsilon \rm{Re}f(\varepsilon)
\times(\frac{1}{k_0+\nu+\varepsilon}-\frac{1}{k_0-\nu-\varepsilon})
\label{sumt2}
\end{equation}
where, $f(\varepsilon)=\phi(\varepsilon, \vec{q})/\sqrt{[\varepsilon/Z(\varepsilon)]^2-\phi^2(\varepsilon, \vec{q})}$. The
minus sign  ahead of Eq.(\ref{sumt2}) comes from the contribution of the
trivial part in Eq.(\ref{approx2}) . The factor $\rm{Re}f(\varepsilon)$ is
obtained because of  the integral of momenta $q$.
\begin{equation}
\int dq\rm{Im}[\Xi(\varepsilon, \vec{q})]=\rm{Im}[\int
dq\Xi(\varepsilon, \vec{q})]=\rm{Im}[-i\pi
f(\varepsilon)]=-\pi\rm{Re f(\varepsilon)}
\end{equation}
We now perform the analytical continuation,  $k_0$ to
$\omega+i\eta$,  and take the imaginary part of the Eqs. (\ref{suml2}),(\ref{sumt2})
\begin{widetext}
\begin{equation}
{\rm{Im}}\int
dqT\sum_{q_0}\Delta_l(K-Q)\Xi(Q)=-{\rm{sign}}(\omega)\pi\int\limits_{0}^{|\omega|}d\varepsilon{\rm{Re}}f(\varepsilon)\rho_{l}^{cut}(|\omega|-\varepsilon, \vec{k}-\vec{q})
\end{equation}
\begin{equation}
-{\rm{Im}}\int
dqT\sum_{q_0}\Delta_t(K-Q)\Xi(Q)={\rm{sign}}(\omega)\pi\int\limits_{0}^{|\omega|}d\varepsilon{\rm{Re}}f(\varepsilon)\rho_{t}^{cut}(|\omega|-\varepsilon, \vec{k}-\vec{q})
\label{img-t}
\end{equation}
\end{widetext}
where, we have made use of Eq. (\ref{gspectral}) and ignored the
$\rho_{l, t}^{pole}$ terms, since the two delta functions can not
be satisfied simultaneously.  It means that the pole terms in the
spectral densities of gluon propagators give no contribution to
the imaginary part of the gap function.  The expressions of
$\rho_{l, t}^{cut}$ have been given in Ref. \cite{RD} in the
limit, $\omega<<p<<m_g$.
\begin{equation}
\rho_{l}^{cut}(\omega, \vec{k}-\vec{q})\simeq\frac{2M^2}{\pi}\frac{\omega}{p}\frac{1}{(p^2+3m_{g}^{2})^2}
\end{equation}
\begin{equation}
\rho_{t}^{cut}(\omega, \vec{k}-\vec{q})\simeq\frac{M^2}{\pi}\frac{\omega
p}{p^6+(M^2\omega)^2}
\end{equation}
where, $M^2\equiv\frac{3\pi}{4}m_{g}^{2}, m_{g}^{2}=\frac{g^2\mu^2}{3\pi^2}$
is the gluon mass of the HDL propagators for two flavor quark
matter at zero temperature.  Now,  we can perform the integral of
the polar angles of $\vec{q}$
\begin{equation}
\int
pdpd\varphi\rho_{l}^{cut}(\omega, \vec{k}-\vec{q})\simeq4M^2\omega\int\limits_{0}^{2\mu}\frac{dp}{(p^2+3m_{g}^{2})^2}\simeq\frac{\sqrt{3}\omega\pi^2}{12m_g}
\end{equation}
\begin{equation}
\int
pdpd\varphi\rho_{t}^{cut}(\omega, \vec{k}-\vec{q})\simeq2M^2\omega\int\limits_{0}^{2\mu}\frac{p^2dp}{p^6+(M^2\omega)^2}\simeq\frac{\pi}{3}
\label{int-cut-t}
\end{equation}
The contribution from the longitudinal gluons to the imaginary
part can be ignored in comparison with that from the transverse
gluons if $\omega<<m_g$,  as is required for the validity of the
approximations of the gluon propagators. Combining Eqs.(\ref{gap}), (\ref{img-t})
,(\ref{int-cut-t}) we obtain

\begin{equation}
\rm{Im}\phi(\omega)=\rm{sign}(\omega)\frac{g^2}{36\pi}\int\limits_{0}^{|\omega|}d\varepsilon\rm{Re}f(\varepsilon)
\label{imag}
\end{equation}

For the sake of completeness, we also include the real part of the
gap function
\begin{equation}
\rm{Re}\phi(\omega)=\frac{g^2}{36\pi^2}\int\limits_{0}^{\omega_0}d\varepsilon
\Big(\ln\frac{\omega_c}{|\omega-\varepsilon|}+\ln\frac{\omega_c}
{|\omega+\varepsilon|}\Big)\rm{Re}f(\varepsilon)
\label{real}
\end{equation}
where $\omega_c=\frac{256\pi^4\mu}{g^5}$ for two quark flavors and $\omega_0\sim m_g$.
The structure of the gap function is consistent with that in
\cite{RD}. For instance , the real part is an even function of the energy while its imaginary
part is an odd function of the energy.

\section{ The imaginary part of the gap }
As in the standard Eliashberg theory\cite{G},  the imaginary part
of the gap function must be derived from  two coupled  equations
of $\rm{Re}\phi(\omega)$ and $\rm{Im}\phi(\omega)$.  However,  it
is important to notice that the forward logaritm in the real part
of the gap equation does not show up in the imaginary part,  which
means that $\rm{Im}\phi$ is down by order $g$ relative to
$\rm{Re}\phi$.  In the leading approximation,  we may ignore
$\rm{Im}\phi$ in Eq. (\ref{real}) and the RHS of Eq. (\ref{imag})
and determine $\rm{Im}\phi$ from the approximate solution for
$\rm{Re}\phi$ in the literature,  i.  e.  the first term of Eq.
(\ref{expansion}).  For this purpose,
we input the real part of the gap function for  our calculations
coming from the results by Sch$\ddot{a}$fer and Wilczek
\cite{TF}. Although their solution is for an imaginary energy , it
actually gives the correct,  up to the subleading order,  results of
the real part of the gap function(see the Appendix),
\begin{equation}
\rm{Re}\phi(\omega)=\left\{
\begin{array}{cc}
\Delta_0, &\rm{if} \hspace{0.2cm}\omega<\Delta_0\\
\\
\Delta_0\sin[\bar{g}\ln\frac{\omega_c}{\omega}]=\Delta_0\cos[\bar g\ln(\frac{|\omega|}{\Delta_0})], &\rm{if}
\hspace{0.2cm}\omega>\Delta_0
\end{array}\right.
\label{Thirtith}
\end{equation}
with
$\Delta_0=2\omega_c\exp(\frac{-3\pi^2}{\sqrt{2}g}-\frac{\pi^2+4}{8})$,
$\bar{g}=g/(3\sqrt{2}\pi)$. By
making use of the Eq.(\ref {Thirtith}),  we obtain the analytic leading order
expressions of the imaginary part of the gap, which is zero for
$0<\omega<\Delta_0$ and is
\begin{equation}
\rm{Im}\phi(\omega)\simeq\bar{g}\Delta_0\frac{\pi}{2}\sin[\bar{g}\ln\frac{\omega}{\Delta_0}]
\label{result}
\end{equation}
for $\omega>\Delta_0$.  Here,  we have  made approximations by ignoring the imaginary part of the gap
in the function $f(\varepsilon)$ in Eq.(\ref{imag}) ,  which means the
energy $q_0$ is on the quasi-particle mass-shell.  So the result
given by Eq. (\ref{result})  corresponds to  Eq. (3.177)  in\cite{philipp}.

The imaginary part  (\ref{imag}) follows from a rigorous analytic
continuation of the gap equation with Euclidean energy.  It is instructive to compare
this result with the analytic continuation of the approximate gap
function in Euclidean space,  which takes the form \cite{D,TF}
\begin{equation}
\phi(i\nu)=\left\{
\begin{array}{cc}
\Delta_0, &\rm{if} \hspace{0.2cm}|\nu|<\Delta_0\\
\\
\Delta_0\sin[\bar {g}\ln(\frac{\omega_c}{|\nu|})], &\rm{if}
\hspace{0.2cm}|\nu|>\Delta_0
\end{array}\right.
\label{Euclidean}
\end{equation}
The analyticity of the quark propagator for sufficiently small $|\omega|$
implies  that of $\phi(\omega)$ under the same condition.  The Euclidean solution
(\ref{Euclidean}) suggests two logarithmic cuts on the real axis,  symmetric with
respect to the imaginary axis and leaving a gap at the origin.  We have then
\begin{equation}
\phi(i\nu)=\Delta_0\sin[\frac{\bar g}{2}(\ln\frac{\omega_c}{0^{+}+i\nu}
+\ln\frac{\omega_c}{0^+-i\nu})]
\end{equation}
for $\nu>>\Delta_0$.  The real part and the imaginary part of its analytic continuation,
$i\nu\to\omega+i0^+$ agree exactly with Eqs.(\ref{Thirtith}),(\ref{result})
for $\omega>>\Delta_0$.

We have also numerically solved the integral Equation  (\ref{imag}) by  using  the
leading order results given by Eq. (\ref{Thirtith}). The energy dependences of the imaginary part of
the gap function at different chemical potentials are depicted in  Fig. 1 and Fig. 2.  We
didn't give the results at higher $\omega$ because the validity of
the approximations for the gluon propagator requires that the
energy $\omega$ can not be too large(see the discussions below the
Eq. (\ref{int-cut-t}) ),  and we choose the maximum value of $\omega$ to be
500MeV.

\begin{figure}
\includegraphics[scale=0.86, clip=true]{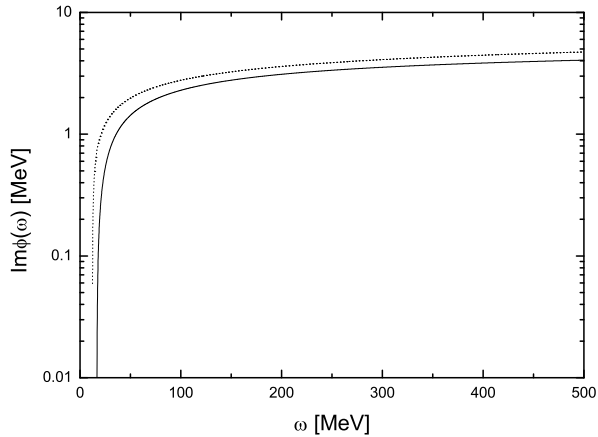}
\caption{\label{fig:epsart}The dependence of $\rm{Im}\phi$ on
$\omega$ at $\mu=500\rm{MeV}$. The solid line shows the approximate
solution and the dashed line shows the numerical solution. }
\end{figure}
\begin{figure}
\includegraphics[scale=0.86, clip=true]{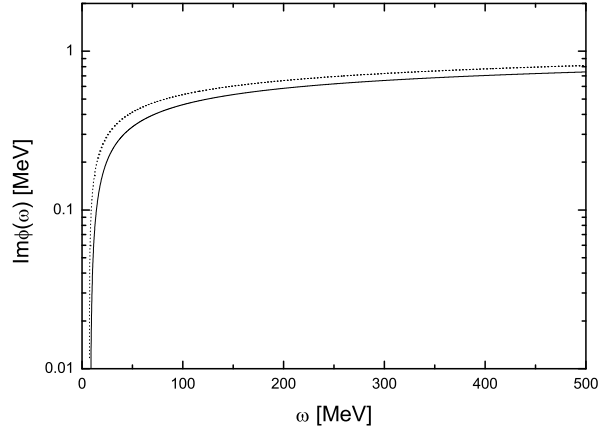}
\caption{\label{fig:epsart} The dependence of $\rm{Im}\phi$ on
$\omega$ at $\mu=5000\rm{MeV}$.  The solid line and the dashed line
show the approximate solution and the numerical solution,
respectively. }
\end{figure}

>From Fig.1 and Fig.2, we know that the contribution from the
off-shell behavior of $q_0$ can be ignored in comparison with that
from the on-shell behavior,  which agrees with the analysis
in\cite{philipp},  i. e.  the off-shell behavior only yields
sufficiently small corrections to $\rm{Im}\phi$.  The imaginary
part of the on-shell gap function near the Fermi surface was
calculated in\cite{philipp} with a different approach,
 The prefactor of the result  Eq.(\ref{result})  agrees with that in\cite{philipp},  which means that the
magnitude of the imaginary part is the same as in\cite{philipp}.

For comparison, We  depict both the approximate analytic result and the numerical result
in Fig. 1 and Fig.  2.  We used the running coupling $g(\mu)$
according to the one-loop beta function as used by
Sch$\ddot{a}$fer and Wilczek \cite{TF}.  The points of intersection
in the abscissa are just the values of $\Delta_0$.  The imaginary
part of the gap function is obviously a  function of $\omega$ and is
zero for $\omega$ smaller than $\Delta_0$ as expected.  At
extremely large densities, i. e.  $\mu=5\rm{GeV}$,  the magnitude of the
imaginary part of the gap function is smaller than one MeV.  For
chemical potential that are of physical interest,
$\mu=500\rm{MeV}$,  the value of the imaginary part can be as large
as several MeV.  We should caution,  however,  in this regime $g$ is
significantly bigger than 1,  and higher order corrections are
probably important.  But to the leading order,  the numerical
results show that the imaginary part of the gap function is down
by $g$  relative to the real part of the gap function,  which is on
the order of 100\rm{MeV} in weak coupling.

A positive imaginary part appears odd since one expect naively that
the complex quasi-particle pole as given by the condition
$\omega^2-(p-\mu)^2-\Delta^2(\omega)=0$ acquires a positive imaginary part for
${\rm Im}\Delta(\omega)<<{\rm Re}\Delta(\omega)$, in violation of the
causality. In this regard, however, one cannot ignore the wave function
renormalization $Z(\omega)$ even at the leading order. The proper
condition for the quasi-particle pole to be used reads $Z^{-2}\omega^2
=(p-\mu)^2+\Delta^2(\omega)$, where
\begin{equation}
Z(\omega)\simeq 1-\bar g^2\Big[\ln\frac{M^2}{\omega^2}+i\pi{\rm sign}(\omega)\Big]
\end{equation}
is the analytic continuation of (\ref{Zfactor})
We have then
\begin{equation}
{\rm Im}\omega\simeq -\pi\bar g|\bar\omega|
+\frac{{\rm Re}\phi(\bar\omega){\rm Im}\phi(\bar\omega)}{\bar\omega}
\end{equation}
with ${\bar\omega}\equiv{\rm Re}\omega$ the solution of
$\bar\omega={\rm sign}(\bar\omega)\sqrt{(p-\mu)^2+[{\rm Re}(\bar\omega)]^2}$.
On writing $x\equiv\bar g\ln\frac{|\bar\omega|}{\Delta_0}>0$, we find
\begin{equation}
{\rm Im}\omega\simeq-\frac{\pi\bar g}{2|\bar\omega|}\Delta_0^2\Big(2\bar ge^{\frac{x}{\bar g}}
-\frac{1}{2}\sin 2x\Big)<0.
\end{equation}
and the quasi-particle pole is indeed below the physical sheet.
It was also suggested in \cite{philipp} that the imaginary part of 
$Z$ is necessary to shift the quasi-particle pole to the complex energy 
plane since the pole in his formulation is on real axis for Z=real even 
with nonzero ${\rm Im}\phi$. We shall come to this difference in the next 
section.

Before ending this section,  we shall sketch the parallel analysis for a
CFL condensate,  which is the most favored with three quark flavors.
Starting with the CFL gap equation with Matsubara energy and following
the same procedure outlined in section 3,  we end up with a pair of
gap equation which amounts to replace the function $f(\varepsilon)$ of
 2SC by \cite{dk}
\begin{equation}
\frac{1}{3}f_1(\varepsilon)+\frac{2}{3}f_2(\varepsilon)
\end{equation}
with
\begin{equation}
f_1(\varepsilon)=\frac{\phi(\varepsilon, \vec{q})}
{\sqrt{[\varepsilon/Z(\varepsilon)]^2-4\phi^2(\varepsilon, \vec{q})}}
\end{equation}
and
\begin{equation}
f_2(\varepsilon)=\frac{\phi(\varepsilon, \vec{q})}
{\sqrt{[\varepsilon/Z(\varepsilon)]^2-\phi^2(\varepsilon, \vec{q})}}.
\end{equation}
The leading order solutions for the real part and the imaginary
part remain given by Eqs.  (\ref{Thirtith}) and (\ref{result}) but
with $\Delta_0$ replaced by $2^{-\frac{1}{3}}\Delta_0$.  Because
of the two gap structure,  one may expect three pieces for the
solution for three different domains, $\omega<\Delta_0$,  $
\Delta_0<\omega<2\Delta_0$ and $2\Delta_0<\omega<\delta$. But the
correction caused by this complication is of higher orders.

\section{Concluding remarks}


In summary we derive perturbatively the equation of the imaginary part of the gap function for
dense QCD in weak coupling and solve this equation numerically. The result show that, up to the
leading order, the imaginary part is small at extremely large densities and can be as large as
several MeV for densities of physical interest. We find that the imaginary part of the gap function
is down by $g$ relative to the real part for the energy $0<\omega<<m_g$,
in agreement with the result of Ref. \cite{philipp} obtained with a different 
approach. In addition we generalize our result from 2SC case to CFL case.

Besides technical differences, the author of \cite{philipp} made the 
assumption that the quasi particle energy function $\epsilon(\phi)$ 
in the function $\Xi$ of (\ref{agreen}) 
depends on the imaginary part of the gap 
according to
\begin{equation}
\epsilon^2(\phi)=(q-\mu)^2+|\phi|^2 = 
(q-\mu)^2+({\rm Re}\phi)^2+({\rm Im}\phi)^2,
\label{phillip}
\end{equation}
generalizing the treatment of \cite{DA} and \cite{RD} where
$\phi$ is valued on the quasi-particle mass shell. But he also raised the 
suspicion about this form because it leads to a real quasi-particle pole with 
a nonzero spetral width \cite{reuter}. 
In our formulation along the line of Ref.\cite{G}, we have 
\begin{equation}
\epsilon^2(\phi)=(q-\mu)^2+\phi^2(\omega)=(q-\mu)^2+[{\rm Re}\phi(\omega)
+i{\rm Im}\phi(\omega)]^2
\label{fhlr}
\end{equation}
with $\phi$ valued off shell, since the form $|\phi|^2$ cannot be maitained in 
an analytic continuation. Although the difference between eqs. (\ref{phillip} 
and (\ref{fhlr}) does not affect the result of ${\rm Im}\phi$ up to the 
order considered in both works, it is conceptual and worth clarifying.
So we did in Sect.II and we conclude that (\ref{fhlr} ) is the right
form of the quasi particle energy with a complex energy-momentum 
dependence if the system conserves both ${\cal P}$ and ${\cal T}$. 
The two NG off-diagonal elements are complex conjugate of each 
other under the same {\it{Matsubara energy}} only (see Eq.(\ref{selfm})). 
For real energy, the complex
conjugation relates the two off-diagonal elements with
{\it{oppisite sign}} of energy (see Eq.(\ref{selfh})). In
addition, the invariance under ${\cal P}$ and ${\cal T}$ renders the
two off-diagonal elements differ by a {\it{constant}} phase factor only
(Theorems 1 and 2). According to (\ref{fhlr}), the imaginary part of the 
gap function does shift the quasi-particle pole off real axis. But it is still 
necessary to have ${\rm Im}Z\neq 0$ to place the quasi-particle pole under 
the branch cut.

 At asymptotically high value of $\mu$, only quarks close to the Fermi surface participate in pairing.
 The analysis carried out here, strictly speaking, applies only to this case with $\omega<<m_g$.
 As $\mu$ is reduced towards the realistic value accessible in a compact star,
 the pairing phase space spread away from the Fermi surface such that the difference
 between $\omega\sim\Delta$ and $\omega\sim m_g$ becomes less clear-cut. The extrapolation of the solution of Ref.\cite{philipp}
  and ours to $\omega\sim m_g$ implies ${\rm Re}\phi\sim{\rm Im}\phi$ there. Therefore, the real and imaginary parts of the gap
   function have to be treated equally and one has to stay with 
Eq.(\ref{fhlr}) for the quasi-particle energy. In addition, the
corrections to the spectral density (\ref{spect-f}) and the
correction from the the imaginary part of $Z(\omega)$ have to be
collected for a more accurate determination of ${\rm
Im}\phi(\omega)$ \cite{philipp}.

At realistic value of $\mu$, the pairing channel is speculated to be dominated by instantons and the color
superconductivity is described by NJL effective action. The magnitude of the gap is expected to be considerably
larger than that of the one-gluon exchange. While the bare four-quark vertex of NJL gives rise to an entirely real gap parameter,
 its one loop correction contains damping terms and may generate an imaginary part of the gap.
 In view of the crudeness of  the weak coupling approximation of NJL, it would be interesting to examine this possibility.

We expect the nonzero imaginary part of the gap will affect transport properties, the interaction rates,
like the neutrino emission rates of color superconductors and thus is useful for understanding the evolution and
structures of a compact star\cite{DTD,alford,AAG,xp, itoh}.

\begin{acknowledgments}
We especially  thank P. Reuter for sending us  his result before sending
to the archive and for his valuable comments. We would like to extend our gratitude to  I.  Giannakis,  D. Rischke
 and  T.  Sch$\ddot{a}$fer for helpful discussions. The work of H. C. R is supported in part by US Department of Energy
under grants DE-FG02-91ER40651-TASKB.   The work of D. F. H.  and H.  C.  R.
 is supported in part by NSFC under grant No.  10575043.  The work of D. F. H.  is
 also  supported in part by  Educational Committee under grants NCET-05-0675 and
704035.  The work of J.R.L.  is supported partly by NSFC under grants 90303007 and 10135030.

\end{acknowledgments}

\appendix
\section{}
In this appendix,  we shall solve the gap equation for ${\rm
Re}\phi(\omega)$ up to the subleading order analytically.
Neglecting the feed back from ${\rm Im}\phi(\omega)$ and the
imaginary part of the wave function renormalization\cite{philipp}.,  the real part
of the gap equation is given by
\begin{widetext}
\begin{equation}
{\rm Re}\phi(\omega)=\frac{\bar g^2}{2}\int_{\Delta_0}^{\omega_0}
d\varepsilon\Big[\ln\frac{\omega_c}{|\omega-\varepsilon|}
+\ln\frac{\omega_c}{(\omega+\varepsilon)}\Big] \frac{{\rm
Re}\phi(\varepsilon)}{\sqrt{Z^{-2}(\varepsilon) \varepsilon^
2-[{\rm Re}\phi(\varepsilon)]^2}} \label{Rephi}
\end{equation}
\end{widetext}
subject to the free boundary condition:
\begin{equation}
Z^{-2}(\Delta_0)\Delta_0^2-[{\rm Re}\phi(\Delta_0)]^2=0.
\label{nonlinear}
\end{equation}

We shall solve the nonlinear gap equation by iterations starting
with a constant gap,   ${\rm Re}^{(0)}\phi(\omega)=\Delta_0$.  In
each step of iterations,  we replace ${\rm Re}\phi(\omega)$ inside
the square root on RHS of (\ref{Rephi}) by that obtained from the
previous step and the integral equation becomes linear subject to
the nonlinear boundary condition (\ref{nonlinear}).  To be more
specific the linear integral equation for the $n-th$ iteration
defines an eigenvalue problem
\begin{equation}
E{\rm Re}\phi^{(n)}(\omega)=\int_{\Delta_0}^{\omega_0} d\varepsilon
K^{(n)}(\omega, \varepsilon){\rm Re}\phi^{(n)}(\varepsilon)
\label{eigen}
\end{equation}
for $\Delta_0<\omega,\varepsilon<\omega_0$ of the kernel,
\begin{widetext}
\begin{equation}
K^{(n)}(\omega, \varepsilon)=\frac{\bar g^2}{2}
\Big[\ln\frac{\omega_c}{|\omega-\varepsilon|}
+\ln\frac{\omega_c}{(\omega+\varepsilon)}\Big]
\frac{1}{\sqrt{Z^{-2}(\varepsilon)\varepsilon^2 -[{\rm
Re}\phi^{(n-1)}(\varepsilon)]^2}} \label{kernel}
\end{equation}
\end{widetext}
with the eigenvalue $E=1$.  The value of
$\Delta_0$ is determined upon identifying $E$ with the maximum
eigenvalue,  $E_{\max}(\Delta_0)$ of $K^{(n)}$,  i. e.
\begin{equation}
E_{\rm max}(\Delta_0)=1.  \label{gapp}
\end{equation}
The normalization of the corresponding eigenfunction,  ${\rm
Re}\phi^{(n)}(\omega)$ is fixed by the equation (\ref{nonlinear})
and the value of ${\rm Re}\phi^{(n)}(\omega)$ for
$0<\omega<\Delta_0$ can be obtained from that for
$\omega>\Delta_0$ by means of the integral equation (\ref{eigen})

The eigenvalue problem (\ref{eigen}) can be analyzed with the
perturbation method developed in\cite{WJH1, WJH2}.  As will be
justified later,  only the first iteration is required for our
purpose.  The kernel of the gap equation for the first iteration
reads
\begin{equation}
K(\omega, \varepsilon)=\frac{\bar g^2}{2}
\Big[\ln\frac{\omega_c}{|\omega-\varepsilon|}
+\ln\frac{\omega_c}{(\omega+\varepsilon)}\Big]
\frac{Z(\varepsilon)}{\sqrt{\varepsilon^2
-Z^2(\varepsilon)\Delta_0^2}},  \label{lead}
\end{equation}
where we have suppressed the superscript of $K(\omega,
\varepsilon)$.  Decompose the kernel according to
\begin{equation}
K(\omega, \varepsilon)=K_0(\omega, \varepsilon)+
K_1^a(\omega, \varepsilon)+K_1^b(\omega, \varepsilon)
+K_1^c(\omega, \varepsilon)+. . .
\end{equation}
where
\begin{equation}
K_0(\omega, \varepsilon)=\bar
g^2\ln\frac{\omega_c}{\omega_>}\frac{1}{\varepsilon},
\end{equation}
with $\omega_>={\rm max}(\omega, \varepsilon)$
\begin{equation}
K_1^a(\omega, \varepsilon) =\bar
g^2\ln\frac{\omega_c}{\omega_>}\frac{Z(\varepsilon)-1}{\varepsilon},
\end{equation}
\begin{equation}
K_1^b(\omega, \varepsilon) =\bar
g^2\ln\frac{\omega_c}{\omega_>}\Big(\frac{1}{\sqrt{\varepsilon^2
-\Delta_0^2}} -\frac{1}{\varepsilon}\Big),
\end{equation}
\begin{equation}
K_1^c(\omega, \varepsilon) =\bar
g^2\Big[\frac{1}{2}\Big(\ln\frac{\omega_c}{|\omega-\varepsilon|}
+\ln\frac{\omega_c}{(\omega+\varepsilon)}\Big)-
\ln\frac{\omega_c}{\omega_>}\Big]\frac{1}{\varepsilon},
\end{equation}
and . . .  represents higher order terms.  We notice two sources for
the logarithm $\ln\frac{\omega_c}{\Delta_0}$ upon integrating the
kernel over $\varepsilon$ with ${\rm Re}\phi(\Delta_0)\neq 0$,
the one corresponds to the forward singularity of the one-gluon
exchange  and the one from the nonvanishing DOS at the Fermi
surface.  Both contribute to $K_0$ and render the eigenvalue of the
order $O(g^2\ln^2\frac{\omega_c}{\Delta_0})$.  Then the condition
(\ref{gapp}) implies that
$\ln\frac{\omega_c}{\Delta_0}\sim\frac{1}{g}$.  Following this rule
and Eq.  (\ref{Zfactor}),  $K_1^a$ is of the order $g$.  Only the
forward logarithm contributes to $K_1^b$ so it is also of the
order $g$.  As to $K_1^c$,  upon a power series expansion of
$\varepsilon$,  both logarithms are removed and its contribution
is beyond the subleading order.  On writing the maximum eigenvalue
as
\begin{equation}
E_{\max}=E_0+E_1^a+E_1^b+. . .
\end{equation}
where $E_0$ is the maximum eigenvalue of $K_0$ and $E_1^a$,
$E_1^b$  the first order perturbation brought about by
$K_1^a$,  $K_1^b$ .  We have
\begin{equation}
E_0=\frac{4\bar g^2}{\pi^2}\ln^2\frac{\omega_c}{\Delta_0},
\end{equation}
\begin{equation}
E_1^a=-\frac{4(\pi^2+4)}{\pi^4}\bar
g^4\ln^3\frac{\omega_c}{\Delta_0},
\end{equation}
\begin{equation}
E_1^b=\frac{2}{\pi^2}\bar g^2\ln 2\ln\frac{\omega_c}{\Delta_0},
\end{equation}
According to (\ref{gap}),  the gap parameter up to the subleading
order reads
\begin{equation}
\Delta_0=\frac{512\pi^4\mu}{g^5}e^{-\frac{3\pi^2}{\sqrt{2}g}-\frac{\pi^2+4}{8}}.
\end{equation}

The zeroth order wave function,  that solves the eigenvalue problem
\begin{equation}
{\rm Re}\phi(\omega)=\bar
g^2\int_{\Delta_0}^{\omega_0}\frac{d\varepsilon}
{\omega^\prime}\ln\frac{\omega_c}{\omega_>}{\rm
Re}\phi(\varepsilon),  \label{leading}
\end{equation}
reads
\begin{equation}
{\rm
Re}\phi(\omega)=A\sin\Big(\bar{g}\ln\frac{\omega_c}{|\omega|}\Big)
=A\cos \Big(\bar{g}\ln\frac{|\omega|}{\Delta_0}\Big),  \label{sol}
\end{equation}
with $A$ a constant to be fixed by the nonlinear boundary
condition (\ref{nonlinear}).  Since the the first order
perturbation of the wave function is suppressed by at least by a
factor $g$ relative to (\ref{sol}),  we may ignore them for the
rest of the construction.  For the same reason the nonlinear
boundary condition can be approximated by $\Delta_0={\rm
Re}\phi(\Delta_0)$.  It follows from the property
\begin{equation}
\frac{d}{d\omega}{\rm
Re}\phi(\omega)\mid_{\omega=\Delta_0^+}\simeq 0
\end{equation}
that $A=\Delta_0$ to the subleading order.  Since the RHS of Eq.
(\ref{eigen}) depends only on ${\rm Re}\phi(\omega)$ for
$\omega>\Delta_0$,  we may apply it to determine the solution for
$0<\omega<\Delta_0$ and the result is ${\rm
Re}\phi(\omega)\simeq\Delta_0$.

The kernel of the gap equation for the next iteration differs from
(\ref{lead}) by the dependence of the gap parameter inside the
square root.  Expanding the cosine of (\ref{sol}),  the additional
perturbation it brought about is of the order of
\begin{equation}
\delta
K(\omega, \varepsilon)\sim-g^2\ln^3\frac{\omega_c}{\Delta_0}
\frac{\Delta_0^2}{(\varepsilon^2-\Delta_0^2)^{\frac{3}{2}}}
\sim -g^3\frac{\Delta_0^2}{(\varepsilon^2-\Delta_0^2)^{\frac{3}{2}}}.
\end{equation}
We have
\begin{equation}
\int_{\Delta_0}^{\omega_0} d\varepsilon\delta
K(\omega, \varepsilon) \sim -g^3\int_1^\infty
dx\frac{\ln^2x}{(x^2-1)^{\frac{3}{2}}}.
\end{equation}
Therefore its contribution to the eigenvalue is beyond the
subleading order.

\end{document}